\documentclass[printer,bibyear]{aa}  
\usepackage{graphicx,psfig,amsmath}
\usepackage{lscape,longtable,afterpage}

\newcommand{\bp}{$\beta$\,Pictoris}

\begin{document}
   \title{Fe\,I in the Beta\,Pictoris circumstellar gas disk}
   \subtitle{II. The time variations in the iron circumstellar gas}
   \author{
   F.~Kiefer\inst{1,2}
   \and  
   A.~Vidal-Madjar\inst{1,2}      
   \and
   A.~Lecavelier des Etangs\inst{1,2}
   \and 
   V.~Bourrier\inst{3}
   \and
   D.~Ehrenreich\inst{3}
   \and
   R.~Ferlet\inst{1,2}
   \and
   G.~H\'ebrard\inst{1,2}
   \and
   P.~A.~Wilson\inst{1,2,4}
 }
   
\authorrunning{Kiefer et al.}

   \offprints{F. Kiefer (\email{flavien.kiefer@iap.fr})}

   \institute{
   CNRS, UMR 7095, 
   Institut d'Astrophysique de Paris, 
   98$^{\rm bis}$ boulevard Arago, F-75014 Paris, France
   \and
   UPMC Univ. Paris 6, UMR 7095, 
   Institut d'Astrophysique de Paris, 
   98$^{\rm bis}$ boulevard Arago, F-75014 Paris, France
   \and
   Observatoire de l'Universit\'e de Gen\`eve, 51 chemin des Maillettes, 1290, Sauverny, Switzerland
   \and
   Department of Physics, University of Warwick, Coventry, CV4 7AL
   }
   
   \date{}
 
  \abstract
{
$\beta$\,Pictoris is a young planetary system surrounded by a debris disk of dust and gas. 
The gas source of this disk could be exocomets (or ``falling and evaporating bodies'', FEBs), 
which produce refractory elements (Mg, Ca, Fe) through sublimation of dust grains at several tens of stellar radii. 
Nearly 1700~high resolution spectra of $\beta$\,Pictoris have been obtained from 2003 to 2017 using the HARPS spectrograph. 
In paper~I, we showed that a very high S/N ratio allows the detection of many weak Fe\,I lines in more than ten excited levels, and we derived the physical characteristics of the 
iron gas in the disk. The measured temperature of the gas ($\sim$1300\,K) suggested that it is produced by evaporation of grains at about 0.3\,au (38\,$R_\star$) from the star.
Here we describe the yearly variations of the column densities of all Fe\,I components (from both ground and excited levels).
The drop in the Fe\,I ground level column density after 2011 coincides with a drop in Fe\,I excited levels column density, as well as in the Ca\,II doublet and a 
ground level Ca\,I line at the same epoch. All drops are compatible together with photoionisation-recombination equilibrium and $\beta$\,Pic like relative abundances, in a 
medium at 1300\,K and at 0.3\,au from $\beta$\,Pictoris. Interestingly, this warm medium does not correlate with the 
numerous exocomets in the circumstellar environnement of this young star.

}

   \keywords{Stars: Individual: Beta Pictoris; Circumstellar disk; Exocomets.}

   \maketitle
%

\section{Introduction}
\label{Introduction}

The \bp\ system is particularly famous for harboring one the most massive debris disk in the close neiborhood of the Sun (Smith \& Terrile 1984). Despite being 20\,Myr old (Mamajek \& Cameron 2014), 
this star has gone through the phase of dissipating dust and gas in its protoplanetary 
disk, which remnant should be a gas-poor dust-poor debris disk. Observations however show that the disk is one of the most dusty gas-poor disk among
known debris disk (Artymowicz et al. 1997, Vidal-Madjar et al. 1998). With the age of \bp\ being much larger than typical timescales of destruction of dust particles ($<$1\,Myr), 
this led to the conjecture that a replenishment from either collisions or evaporation of hidden planetesimals was necessary to explain the overabundance of dust in this system 
(Backman \& Paresce 1993, Lecavelier et al. 1996). 

Moreover, absorption spectroscopy of $\beta$\,Pic revealed the presence of a stable gas component at the star's radial velocity
and variable absorptions attributed to transiting star-grazing exocomets (Ferlet et al. 1987, Beust et al. 1990) ; see also the review of Vidal-Madjar et al. (1998). The origin of the stable gas 
disk is still unknown, although highly suspected to be connected to the star-grazing exocomets that strongly evaporate dust and ionized species.
Exocomets might be the common vector of gas and dust replenishment in the disk of $\beta$\,Pictoris (Weissman 1984, Lecavelier et al. 1996, Li et al. 1998).
 
\bp\ was observed with the HARPS instrument (see {\it e.g.} Pepe et al.\ 2011) for several years, from 2003 to 2017 thanks to large programs of follow-up (Lagrange et al. 2012, Kiefer et al. 2014, Lagrange et al. 2018). 
Thousands of spectra were gathered and several hundreds of exocometary events detected (Kiefer et al. 2014). 
In Paper~1~(Vidal-Madjar et al. 2017, VM17 hereafter), stacking the 1686 HARPS spectra collected between 2003 and 2015, we presented the measurements of the physical properties 
of the iron gas from the detection of a large number of FeI absorption lines, from the ground level up to the 12969~cm$^{-1}$ excited level. 
We concluded that the measured temperature of the gas ($\sim$1300\,K) coincides with the sublimation temperature of iron from solid compounds, and therefore that this gas is 
likely produced by the evaporation of grains at about 38\,$R_\star$ or 0.32\,au from the star.
Moreover, the ground level absorption lines of Fe\,I presented evidences of two components at different radial velocities. The component centred on the same radial velocity as the one found for the 
excited levels at 20.41$\pm$0.05\,km\,s$^{-1}$ for which we derived the temperature of 1300\,K, and a blueshifted component at 20.07$\pm$0.02\,km\,s$^{-1}$.

Independently, Welsh \& Montgomery (2016) found that the ground level Fe\,I absorption lines in the same $\beta$\,Pic HARPS spectra were experiencing a strong drop of equivalent width between 2011 and 2013. 
They proposed that the source of the Fe\,I experiencing the drop is the "D-family" of exocomets identified in Kiefer et al. (2014). These exocomets are composed of strongly evaporating 
bodies lying about a common orbit, with heliocentric radial velocities in the range +20 to +50\,km\,s$^{-1}$. 

Here we explore in more details the time variations of both ground and excited levels absorptions of Fe\,I in the "stable" circumstellar medium, and how they correlate with exocomet 
activity in the vicinity of the star. 

The analysis is presented in Sect.~3. The discussions and conclusion are given in Sect.~4 and 5, including a study on the variations also observed 
in the Ca\,II doublet (at 3934 and 3968\,\AA) and the Ca\,I line (at 4227\,\AA).

\section{Observations}
\label{Observations}

\begin{figure}
\begin{minipage}[b]{\columnwidth}	  
\includegraphics[angle=90,trim=0.cm 0.cm 0.0cm 0.5cm,angle=-90,clip=true,width=0.95\columnwidth]{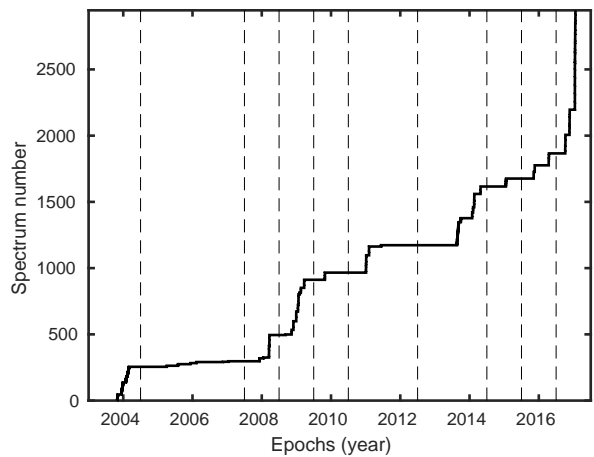}	
\hspace{0.36\textwidth}\\
\end{minipage}
\caption[]{
The time distribution of the HARPS observing nights used in our study. 
The horizontal dotted lines show the limits of the selected observing periods.
}
\label{repartition}
\end{figure}

\bp\ has been observed with the HARPS spectrograph mounted on the 3.6m telescope of La Silla (ESO Chile) 
from 2003 to 2017 on a (mostly) regular basis (except between 2004 and 2007 and in 2012). 
Since \bp\ is observable only during summer in southern hemisphere, the observations were done 
essentially from September to April of each season. The 1D-spectra were extracted via the standard, most recent, 
HARPS pipeline (DRS 3.5) including localization of the spectral orders on the 2D-images, optimal order extraction, 
cosmic-ray rejection, wavelength calibration, flat-field corrections, and 1D-reconnection of the spectral orders 
after correction for the blaze. More details can be found in VM17. 
We organized the spectra into different samples, each constituting one summer 
of observation, with the exception of the period 2004-2007. During this period, the average number of spectra observed 
per summer was 14, which is very small compared to the other periods average (234 spectra per summer). 
For that reason we created a special sample covering the 3 summers from 2004 to 2007. 
This repartition is summarized in Table~\ref{tab:repartition} and Fig.~\ref{repartition}. 

\begin{table}
\caption{\label{tab:repartition} 
The repartition of the 2946 HARPS observations. The signal-to-noise is calculated next to the main Fe I line between 3859.30\,\AA~and 3859.70\,\AA.}
\begin{tabular}{@{}crrccc@{}}
\hline
Periods    &   $N_\text{obs}$  &  $N_\text{night}$ & $\Delta t_\text{span}$    &       SNR    &  Average   \\ 
 (year)    &                   &                   & (day)     &   (3860\,\AA) &   Epoch     \\
\hline
2003-2004  & 255               &      53           &  127      &  962      &    2004.025     \\
2004-2007  & 42                &      14           &  651      &  390      &    2006.235     \\
2007-2008  & 198               &      12           &  108      &  492      &    2008.203     \\
2008-2009  & 417               &      12           &  196      &  909      &    2009.070     \\
2009-2010  & 54                &      1            &  1        &  278      &    2009.847     \\
2010-2011  & 207               &      4            &  159      &  467      &    2011.081     \\
2013-2014  & 453               &      19           &  250      &  799      &    2013.929     \\
2014-2015  & 60                &      4            &  4        &  493      &    2015.091     \\
2015-2016  & 190               &      4            &  162      &  591      &    2016.063     \\
2016-2017  & 1080              &      11           &  109      &  639      &    2017.041     \\
\hline
\end{tabular}
\end{table}

Because of the high S/N ratio achieved, we were able to detect numerous Fe\,I absorption lines of the ground level, as well as, of several excited levels. 
All the detected Fe\,I lines are listed in Table~\ref{observations}. The energy of the initial 
electronic levels of the corresponding transitions are summarized in Table~\ref{tab:sum_levels}. In the paper we always refer to the individual levels by 
their energy expressed in cm$^{-1}$, as e.g. Fe\,I$_{416}$ for the first excited level at 416\,cm$^{-1}$. We will collectively refer to the excited levels as 
Fe\,I$_\text{Exc}$, as opposed to the ground level Fe\,I$_0$. 

The profile of a few individual lines can be found in paper~I, where we also describe the fitting procedure to derive the column densities, velocities and turbulent parameter. 

\begin{table}
\centering
\caption{\label{tab:FeIlines} 
Transition parameters for the detected FeI lines, ($i$ lower level, $k$ upper level.  $A$ is the 
transition probability in s$^{-1}$, $f$ is the oscillator strength, and $E$ is the energy in cm$^{-1}$.
}
\begin{tabular}{ccccc}
 \hline
 Fe\,I lines   & A$_{\rm ki}$  & f$_{\rm ik}$ & E$_{\rm i}$ & E$_{\rm k}$ \\
  (\AA ) & (s$^{-1}$) &  & (cm$^{-1}$) & (cm$^{-1}$) \\ 
 \hline
3795.002 & 1.15$\times$10$^7$ & 3.47$\times$10$^{-2}$ & 7986 & 34329  \\
3799.547 & 7.31$\times$10$^6$ & 2.04$\times$10$^{-2}$ & 7728 & 34040  \\
3812.964 & 7.91$\times$10$^6$ & 1.23$\times$10$^{-2}$ & 7728 & 33947  \\
3815.840 & 1.12$\times$10$^8$ & 1.90$\times$10$^{-1}$ & 11976 & 38175  \\ 
3820.425 & 6.67$\times$10$^7$ & 1.20$\times$10$^{-1}$ & 6928 & 33096  \\
3824.443 & 2.83$\times$10$^6$ & 4.83$\times$10$^{-3}$ &    0 & 26140  \\
3825.881 & 5.97$\times$10$^7$ & 1.02$\times$10$^{-1}$ & 7377 & 33507  \\
3827.822 & 1.05$\times$10$^8$ & 1.65$\times$10$^{-1}$ & 12561 & 38678  \\
3834.222 & 4.52$\times$10$^7$ & 7.13$\times$10$^{-2}$ & 7728 & 33802  \\
3840.437 & 4.70$\times$10$^7$ & 6.24$\times$10$^{-2}$ & 7986 & 34017  \\
3841.047 & 1.36$\times$10$^8$ & 1.80$\times$10$^{-1}$ & 12969 & 38996  \\ 
3849.966 & 6.05$\times$10$^7$ & 4.49$\times$10$^{-2}$ & 8155 & 34122  \\
3856.371 & 4.64$\times$10$^6$ & 7.39$\times$10$^{-3}$ &  416 & 26340  \\
3859.911 & 9.69$\times$10$^6$ & 2.17$\times$10$^{-2}$ &    0 & 25900  \\
3865.523 & 1.55$\times$10$^7$ & 3.47$\times$10$^{-2}$ & 8155 & 34017  \\
3878.573 & 6.17$\times$10$^6$ & 8.36$\times$10$^{-3}$ &  704 & 26479  \\
3886.282 & 5.29$\times$10$^6$ & 1.20$\times$10$^{-2}$ &  416 & 26140  \\
3895.656 & 9.39$\times$10$^6$ & 7.13$\times$10$^{-3}$ &  888 & 26550  \\
3899.707 & 2.58$\times$10$^6$ & 5.89$\times$10$^{-3}$ &  704 & 26340  \\
3906.479 & 8.32$\times$10$^5$ & 1.90$\times$10$^{-3}$ &  888 & 26479  \\
3920.257 & 2.60$\times$10$^6$ & 1.79$\times$10$^{-2}$ &  978 & 26479  \\
3922.911 & 1.08$\times$10$^6$ & 3.19$\times$10$^{-3}$ &  416 & 25900  \\
3927.919 & 2.60$\times$10$^6$ & 1.00$\times$10$^{-2}$ &  888 & 26340  \\
3930.296 & 1.99$\times$10$^6$ & 6.46$\times$10$^{-3}$ &  704 & 26140  \\
4045.812 & 8.62$\times$10$^7$ & 2.12$\times$10$^{-1}$ & 11976 & 36686  \\
4063.594 & 6.65$\times$10$^7$ & 1.65$\times$10$^{-1}$ & 12561 & 37163  \\
4071.738 & 7.64$\times$10$^7$ & 1.90$\times$10$^{-1}$ & 12969 & 37521  \\
4271.760 & 2.28$\times$10$^7$ & 7.62$\times$10$^{-2}$ & 11976 & 35379  \\ 
4307.902 & 3.38$\times$10$^7$ & 1.21$\times$10$^{-1}$ & 12561 & 35768  \\
4325.762 & 5.16$\times$10$^7$ & 2.03$\times$10$^{-1}$ & 12969 & 36079  \\
4383.544 & 5.00$\times$10$^7$ & 1.76$\times$10$^{-1}$ & 11976 & 34782  \\
4404.750 & 2.75$\times$10$^7$ & 1.03$\times$10$^{-1}$ & 12561 & 35257  \\
 \hline
\end{tabular}
\label{observations}
\end{table}

\begin{table}
\centering
\caption{\label{tab:sum_levels}Fe\,I initial levels of the 32 electronic transitions detected in absorption in the HARPS spectrum of $\beta$\,Pictoris.}
\begin{tabular}{rrrr}
 \hline
Energy level   & Energy level & J  & Number   \\
(cm$^{-1}$)   &  (K)  &  & of lines  \\
 \hline
0  &  0  & 4 & 2  \\
416  & 598 & 3 & 3  \\
704 & 1013 & 2 & 3  \\
888 & 1278 & 1 & 3  \\
978 & 1407 & 0 & 1  \\
6928 & 9968 & 5 & 1  \\
7377 & 10614 & 4 & 1  \\
7728 & 11120 & 3 & 3  \\
7986 & 11490 & 2 & 2  \\
8155 & 11733 & 1 & 2  \\
11976 & 17232 & 4 & 4  \\
12561 & 18073 & 3 & 4  \\
12969 & 18660 & 2 & 3  \\
\hline
\end{tabular}
\label{levels}
\end{table}

\begin{figure*}
\begin{minipage}[b]{\textwidth}	  
\includegraphics[trim=0.cm 0.4cm 0.cm 0.75cm,angle=0,clip=true,width=0.34\textwidth]{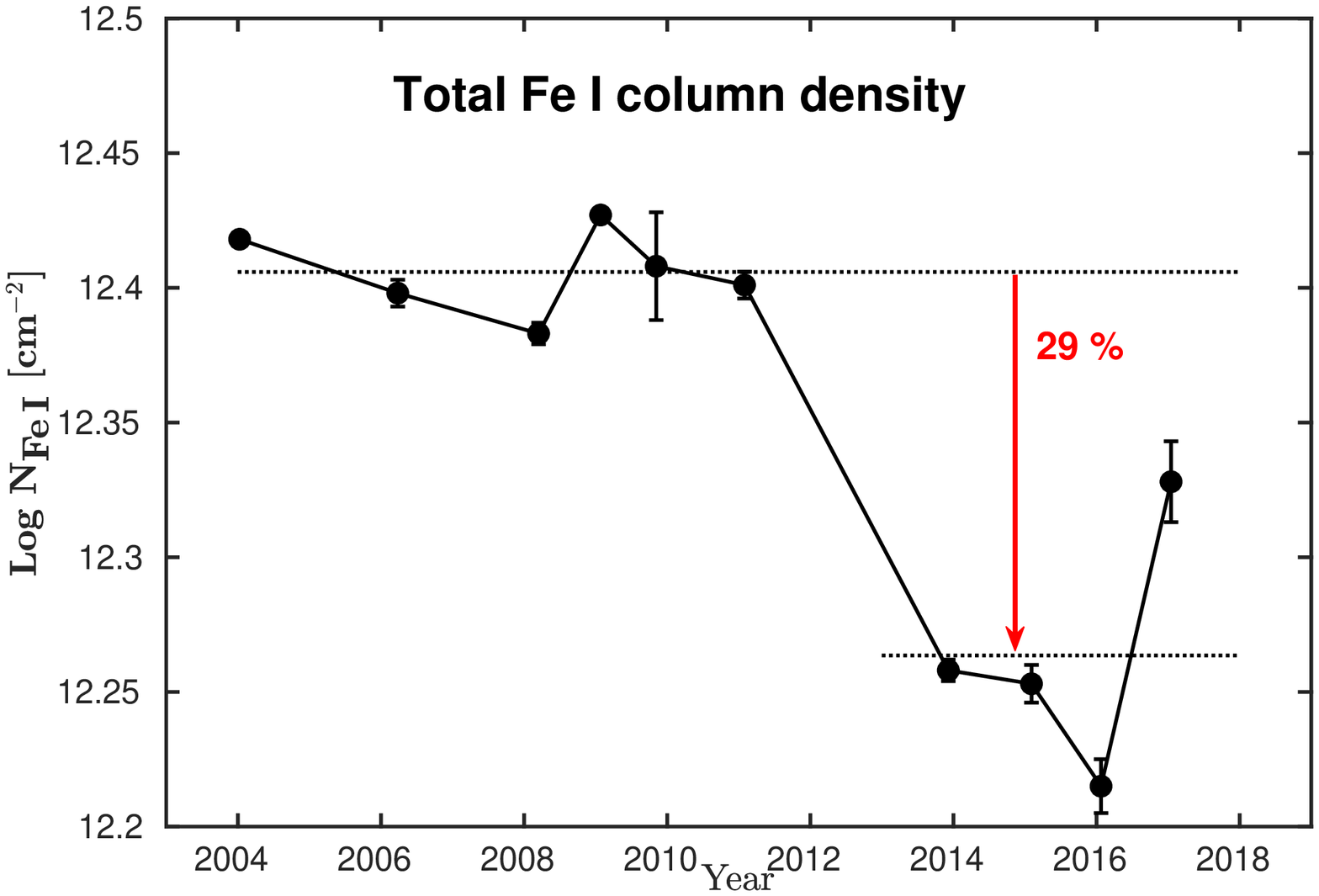}	
\includegraphics[trim=0.cm 0.4cm 0.cm 0.75cm,angle=0,clip=true,width=0.33\textwidth]{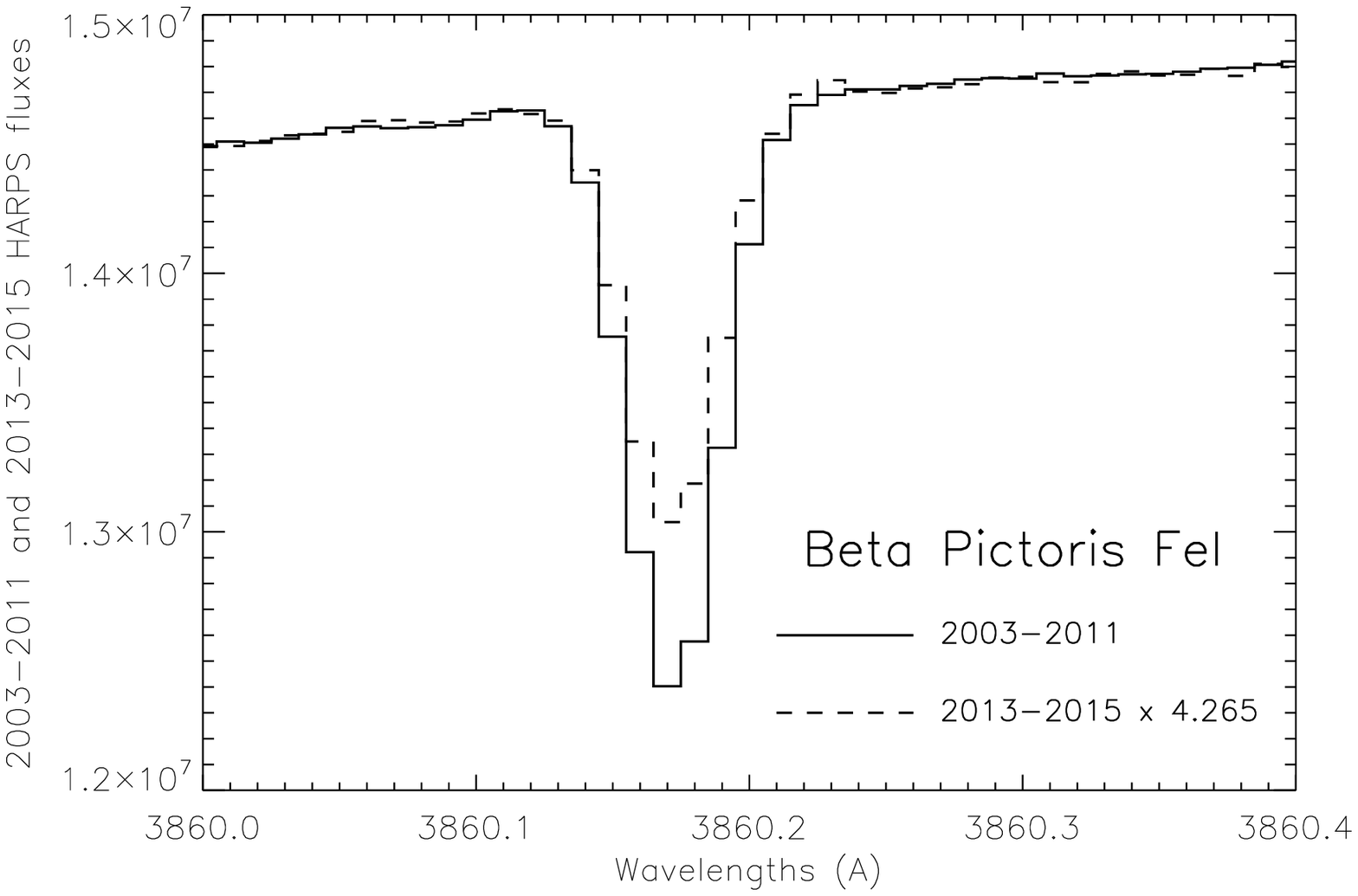}	
\includegraphics[trim=0.cm 0.4cm 0.cm 0.75cm,angle=0,clip=true,width=0.33\textwidth]{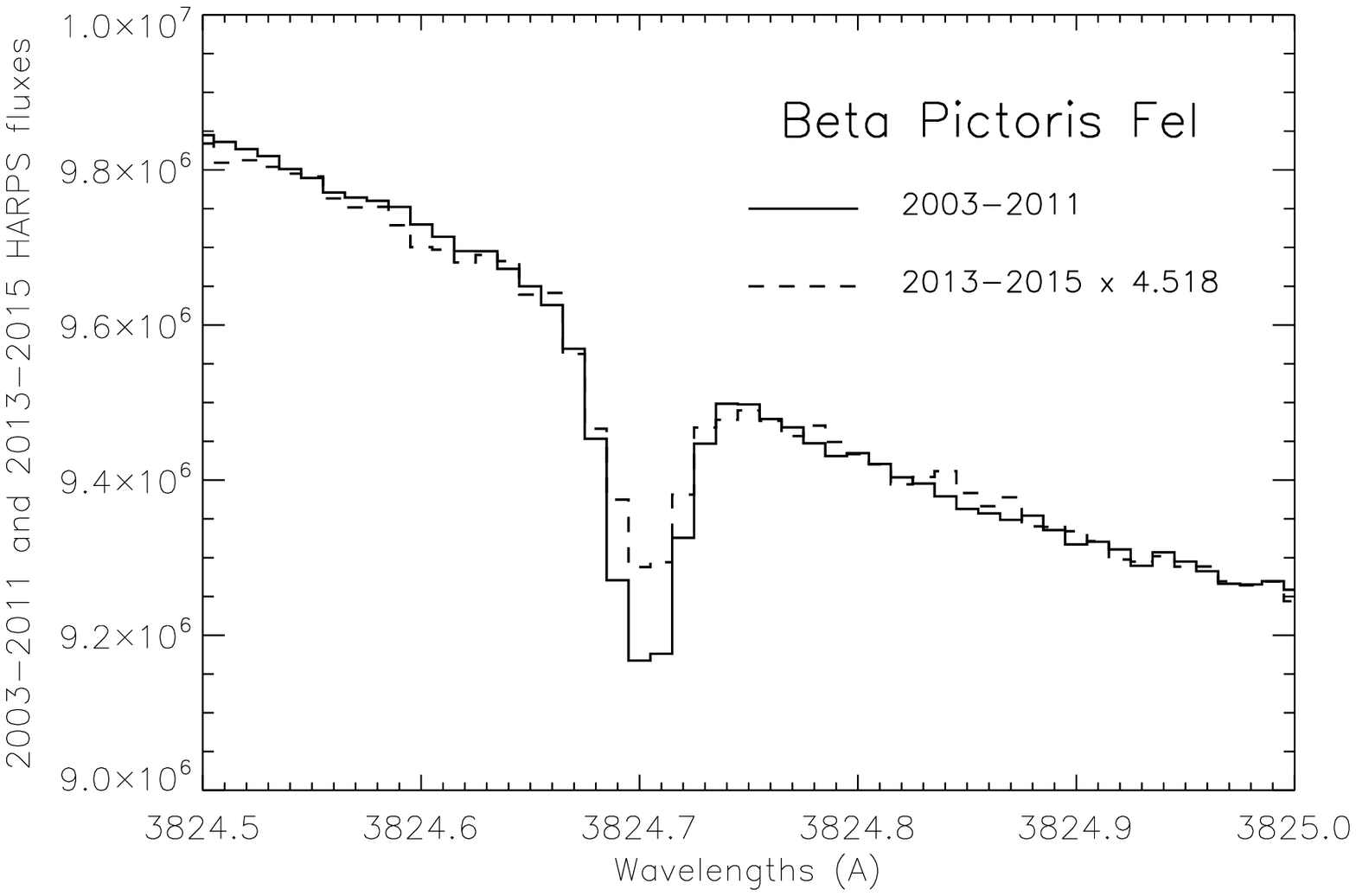}	
\hspace{0.36\textwidth}\\
\end{minipage}
\caption{\label{FeI_temporal_variations}Column density variations in the ground state of the circumstellar Fe\,I of $\beta$\,Pic. \textbf{Left:} Temporal variations of the total Fe\,I$_0$ column density showing a 30\,\% drop after year 2011.
\textbf{Middle and right:} Sum of the Fe\,I HARPS fluxes, over the two separate periods, from 2003 to 2011 (solid) and after 2011 (dashed) at the ground state Fe\,I transition lines wavelengths, 
3860~\AA (middle) and 3824\,\AA (right).}
\end{figure*}

\section{Temporal variations of Fe\,I circumstellar gas} 

\subsection{Variation of the Fe\,I ground level} 
\label{sec:ground}

The time variation of the ``stable'' Fe\,I ground level absorption lines is given in Table~\ref{tab:FeIvariation}, and plotted on Fig.~\ref{FeI_temporal_variations}. We fitted one single 
component for each separated period of observation, using the \verb+Owens.f+ code~(H\'ebrard et al. 2002, Lemoine et al. 2002). It fits a Gaussian line-spred-function of 3.6 pixels 
wide, further broadened by a turbulent parameter $b$, centered on a radial velocity $v$ and with depth fixed by the column density $N$ and the levels transition parameters as 
given in Table~\ref{tab:FeIlines}. The continuum is fitted simultaneously using a 4-degree polynomial. The two Fe\,I lines of the ground level are plotted on Fig.~\ref{FeI_temporal_variations}. 

\begin{table}
\caption{Properties of the FeI gas over the different epochs of observations, evaluated by fitting simultaneously the two absorption lines from the ground state 
(3859.30\,\AA~and 3859.70\,\AA) with a single component.}
\label{tab:FeIvariation}
\begin{tabular}{@{}ccccc@{}}
\hline
Periods   & Nb.  &   v            & b              & $\log$N$_{\rm FeI}$  \\ 
 (year)   & Obs. & (km/s)         & (km/s)         &    (cm$^{-2}$)      \\
\hline
2003-2004 & 215  & 20.15$\pm$0.01 &	0.73$\pm$0.02 &	12.418$\pm$0.002 \\
2004-2007 & 42   & 20.16$\pm$0.02 &	0.81$\pm$0.06 &	12.398$\pm$0.005 \\
2007-2008 & 198  & 20.20$\pm$0.01 &	0.84$\pm$0.04 &	12.383$\pm$0.004 \\
2008-2009 & 417  & 20.21$\pm$0.01 &	0.77$\pm$0.02 &	12.427$\pm$0.002 \\
2009-2010 & 54   & 20.22$\pm$0.02 &	0.58$\pm$0.13 &	12.408$\pm$0.020 \\
2010-2011 & 207  & 20.24$\pm$0.02 &	0.93$\pm$0.05 &	12.401$\pm$0.005 \\
2013-2014 & 453  & 20.14$\pm$0.01 &	0.87$\pm$0.03 &	12.258$\pm$0.004 \\
2014-2015 & 60   & 20.18$\pm$0.02 &	0.85$\pm$0.07 &	12.253$\pm$0.007 \\
2015-2016 & 190  & 20.16$\pm$0.02 &	0.47$\pm$0.08 &	12.215$\pm$0.010 \\
2016-2017 & 1080 & 20.20$\pm$0.01 &	0.40$\pm$0.05 &	12.328$\pm$0.015 \\
\hline
\end{tabular}
\end{table}

The column density changes significantly, showing first a stable value of the order of 2.5~10$^{12}$~cm$^{-2}$ lasting about 7 years, followed by a 3.5$\sigma$ drop 
down to about 1.8~10$^{12}$~cm$^{-2}$, lasting at least 4 years. This drop of about 30~\%\ in the Fe\,I total column density was already noted by Welsh \& Montgomery (2016) from the same 
HARPS observations. It seems that the column density is increasing again in 2017. Future observations should confirm this recent evolution.

The turbulent parameter $b$ is stable up to a few hundreds of m\,s$^{-1}$ indicating this gas ensemble is thermodynamicaly stable over nearly 12 years of observation.  
However, we observe a slow and monotonic reddening of the ground level absorption over years up until 2011, followed by a sudden blueshift simultaneous to the column density drop. Thus the column density variation must be stronger in the red wing than on the blue wing of the Fe\,I line. 

In the light of the two components model of VM17, it strongly 
suggests that only the reddest component at 20.4\,km\,s$^{-1}$ is experiencing the drop. 
Moreover, we show in the next section that a column density drop is also observed in the absorption lines of the excited levels that are centered on 20.4\,km\,s$^{-1}$. 
We will revisit the Fe\,I ground level absorption lines analysis with a 2 components model in Section~\ref{sec:rev-ground}.

\vspace{-0.5cm}
\subsection{Variation of the Fe\,I excited levels} 
\label{sec:excited}

To further study the time behaviour of the Fe\,I source, we examine the variations of the Fe\,I excited levels absorption lines. These evaluations are given in Table~\ref{tab:FeI_exc_variation}, 
as calculated by fitting simultaneously all Fe\,I excited levels absorption lines with \verb+Owens.f+. We fixed that the radial velocity and turbulent parameters $v_\text{Fe\,I Exc}$ and $b_\text{Fe\,I Exc}$
are common to all excited levels, while the column density can vary separately for each individual level. Only the strongest 
transitions show well distinguishable absorptions, since the noise wipes the weaker transition signatures away.

In VM17, we have observed for the whole 2003-2015 period that $v_\text{Fe\,I Exc}$=20.41$^{+0.03}_{-0.05}$\,km\,s$^{-1}$, and $b_\text{Fe\,I Exc}$=1.01$\pm$0.06\,km\,s$^{-1}$. From 
Table~\ref{tab:FeI_exc_variation} we can see that in accordance to VM17, both the $v$ and $b$ of excited levels component remain stable in time about 20.39\,km\,s$^{-1}$  and 
0.96\,km\,s$^{-1}$ respectively, except after 2014 where the velocity is more poorly determined. Conversely, the column densities from the excited levels present a time behaviour 
similar to the ground level with an important drop after 2011. 

\begin{figure}[hbt]\centering
\includegraphics[width=84mm]{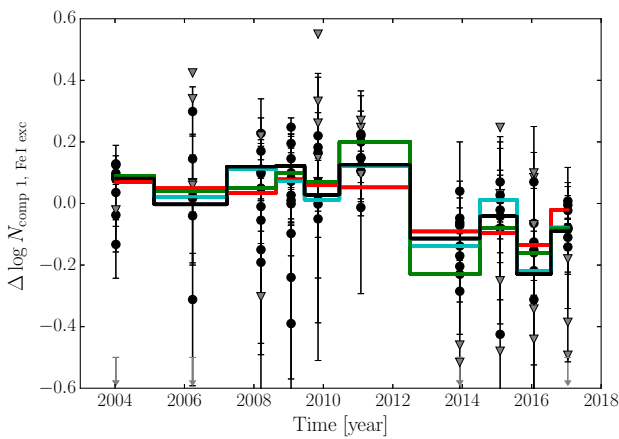}
\caption[]{The spread and evolution of the (log) Fe I column density among the 12 excited levels over time. The black circles refer to each excited level’s column density 
relative to its median value as observed from 2003 to 2017. The grey triangles are upper limits derived from weaker absorptions. A few points outside the 
plot window are figured by arrows next to the axis.  
The black solid line shows how the total column density of excited Fe I varies over time. The cyan and green solid lines 
show the evolution of the column densities derived respectively for the Fe\,I$_{416}$ and Fe\,I$_{6928}$ excited levels. 
The Fe\,I$_0$ variations derived in Table~\ref{tab:FeIvariation} are shown as a solid red line. 
}
\label{Temporal_variations2}
\end{figure}

To have a better look at the column density variations of the excited levels, we plotted in Fig.~\ref{Temporal_variations2} all excited levels values (and upper limits), compared to the 
ground level N$^{\rm Total}_{\rm FeI~0}$ variations.  The cleanest Fe\,I$_{416}$ and Fe\,I$_{6928}$ levels and the total column density in all excited levels are highlighted. All levels 
column densities are normalized to their corresponding median value before the drop, over the 2003 - 2011 period. To calculate the total column density and their uncertainties, we 
drew at each epoch 10$^4$ series of estimations of the column density at all excited levels according to their measurements distribution, then calculated the sum leading to 10$^4$ 
estimations of the total column density, and determined the final measurement as the median of the sample with errorbars as the 1-$\sigma$ percentiles.

The drop observed on the total column density of the excited levels is significant at the 6-$\sigma$ level, with an amplitude of 37$\pm$4\%. It is larger for the excited levels than for 
the ground level, for which we measure a 31$\pm$2\% relative decrease in the column density, leading to a 2-$\sigma$ significant difference in amplitude. 
This suggests again that only part of the ground level is undergoing the same drop than the excited levels. 

\subsection{Revisiting the ground level variations with a 2 components model}
\label{sec:rev-ground}

It has been proposed by VM17 that the absorption lines of the ground level can be divided into 2 components, one of which, at about 20.4\,km\,s$^{-1}$, is common with excited 
levels transition lines, and a second one lying at about 20.0\,km\,s$^{1}$. In the rest of the paper, we will refer to these 2 hypothesized components as the \textit{red component} 
and the \textit{blue component} respectively. Besides the variations observed in the excited levels, additional clues suggest two components might be indeed present 
in the Fe\,I ground level lines. 

First, as observed in section~\ref{sec:ground}, the velocity variations of the one-component fit of the Fe\,I$_0$ absorption lines show that simultaneously to the column density drop there is a 
blueward shift of the centroid wavelength, suggesting that only the red part of the lines undergoes a drop. Second, the amplitude of the drop in ground level column density is 
2-sigma lower than the excited levels column density drop suggesting that only a part of the total amount of Fe\,I in ground level is vanishing. 
Revisiting the analysis of the Fe\,I$_0$ absorption lines using a 2 components model should confirm these assertions.

From the fit of the excited levels absorption lines in Table~\ref{tab:FeI_exc_variation}, the radial velocity of the red component fluctuates about the average value $v_2$$=$$20.39$\,km\,s$^{-1}$,
without any trend. We will thus assume that the red component radial velocity is constant and fix it to this value. We then proceed to a 2-components fit of the ground level absorption line, 
 adding a new component and letting all parameters free except $v_2$. The results are given in Table~\ref{tab:ground_2comp}.

We find that the derived radial velocity of the blue component is weakly fluctuating around the average value $v_1$=19.77\,km\,s$^{-1}$. However, the derived turbulence parameters 
have strong scatters with values lower than 0.05\,km\,s$^{-1}$ for $b_1$ and as high as 1.5\,km\,s$^{-1}$ for $b_2$. This is not surprising since we are fitting strongly blended features.  
Other parameters should be fixed to lower such overfitting effects. 

We thus fix both radial velocities $v_1$ and $v_2$ to respectively 19.77\,km\,s$^{-1}$ and 20.39\,km\,s$^{-1}$. This time the values of the turbulent parameters are better 
behaved, with $b_1$$\sim $$0.11$$\pm $$0.02$\,km\,s$^{-1}$  and $b_2$$\sim $$0.77$$\pm $$0.09$\,km\,s$^{-1}$. The two periods 2015-2016 and 2016-2017 stands aside: 
as for the excited levels the transition line is not well fit with the common model (for excited levels we measured $v$$\sim $$20.20$\,km\,s$^{-1}$). This shift is likely to be imputed 
to the upgrade of the instrument done in June 2015\footnote{https://www.eso.org/sci/publications/messenger/archive/no.162-dec15/messenger-no162-9-15.pdf.}. 
Fortunately, it does not impact the measurement of the column density, so the 2015-2017 period remains valuable for the present analysis.

\begin{figure}
\includegraphics[width=89mm]{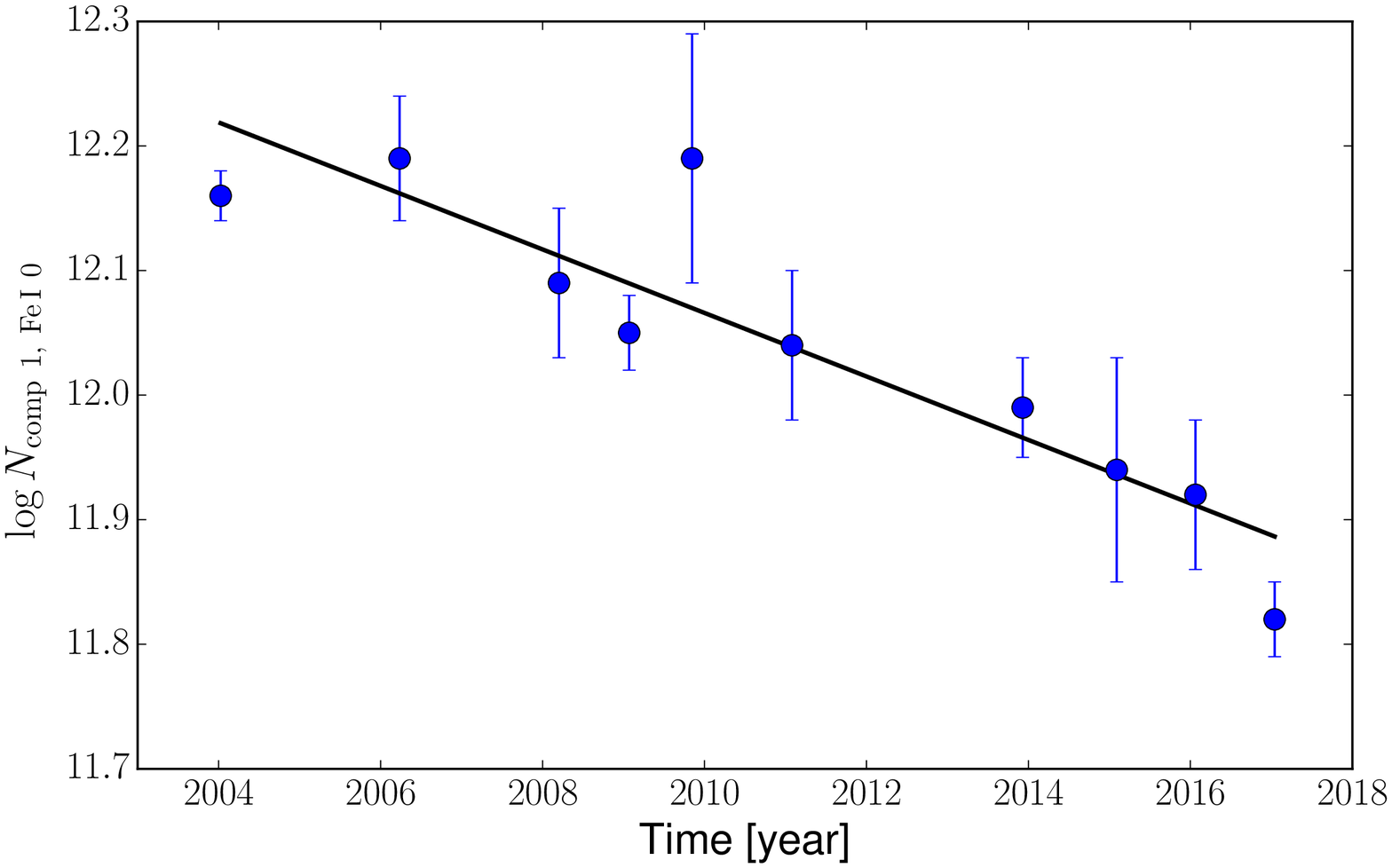}
\includegraphics[width=89mm]{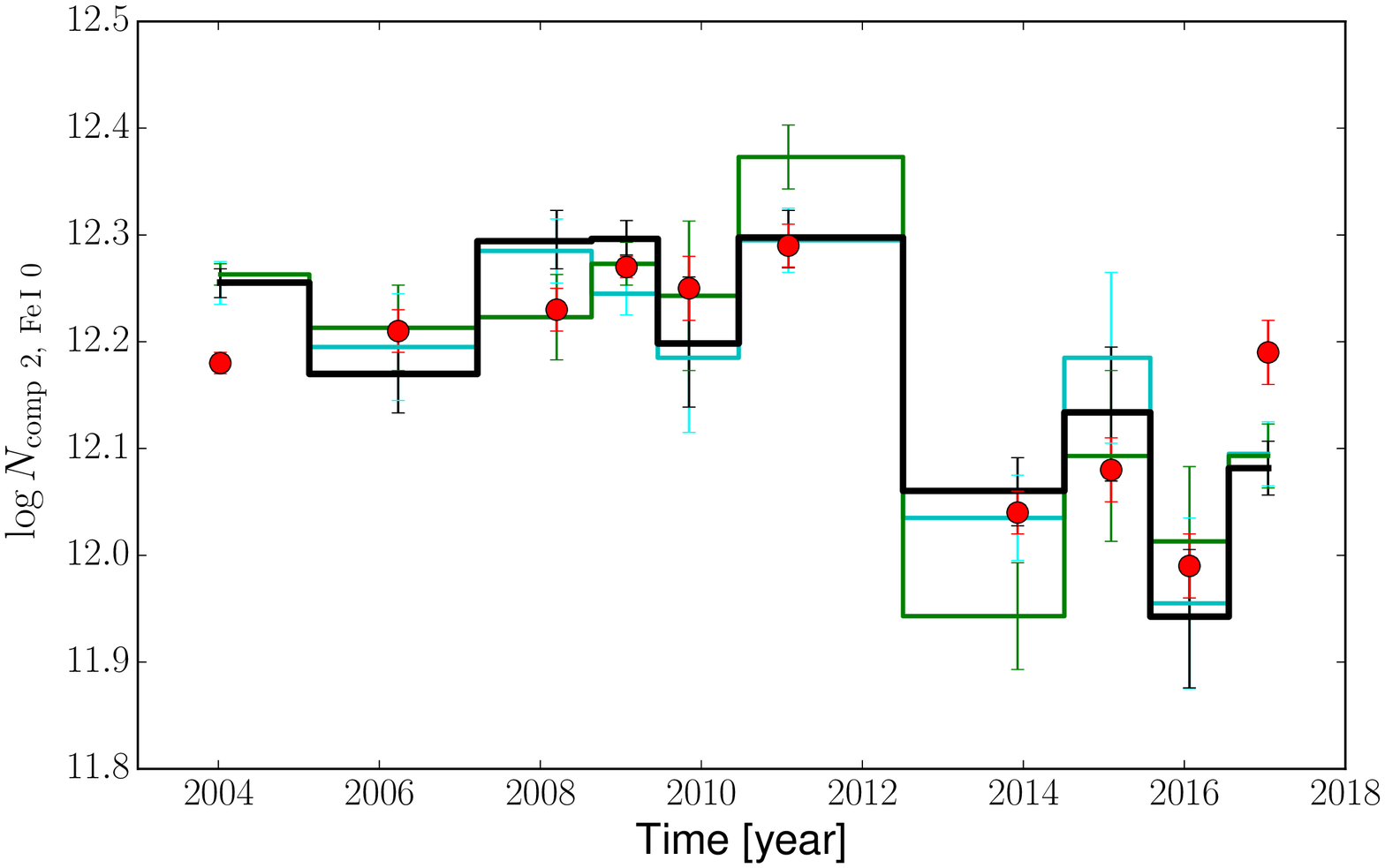}
\caption{\label{fig:ground_2c}Column density variations measured in the Fe\,I ground level absorption lines. Top panel: blue component at 19.80\,km/s. The solid black line 
is a linear model fit of the data. Bottom panel: red component at 20.4 km/s. The coloured solid lines are the variations observed in excited levels column 
densities of Fe\,I$_{416}$ (cyan), Fe\,I$_{6928}$ (green) and total excited Fe\,I (black), as given in Table~\ref{tab:FeI_exc_variation}. They are scaled up to ground 
level column density.}
\end{figure}

The measured ground level column density variations in each component are plotted in Fig.~\ref{fig:ground_2c}. The column density of the blue 
component is found monotonically decreasing, while the red component undergoes a strong drop. This difference is especially pronounced between 2011 and 2013 during the drop,
with the red component column density divided by a factor of 2, and the blue component remaining essentially stable. 
Interestingly, the red variations follow well the excited levels variarions with a slight increase of column density between 2003 and 2011. 

A Pearson's R test shows that the total excited levels column density calculated in Table~\ref{tab:FeI_exc_variation} explains the variations of the red component with $R$=0.87 
(null-hypothesis rejection probability $p$=9.8\,10$^{-4}$), while it only explains the variations of the blue component with $R$=0.59 ($p$=0.074). This means that the red 
component measurements, having also smaller errorbars ($\sim$0.02) than the blue component column density ($\sim$0.06), are much better explained by the excited levels 
variations. However it is difficult to firmly conclude on the compatibility of the blue component column density variations with the excited levels, because although they have larger 
scatter, they also have larger measurement uncertainties.

Conducting another Pearson's R test on linear relationship of the components column density with time shows that the blue component is better explained by a 
continuous decrease through the different epochs, with $R$=-0.90 ($p$=3.7\,10$^{-4}$), while the red component has only $R$=-0.56 ($p$=0.093). 

The red and the blue are thus most likely uncorrelated, and the drop in the ground level column density is better explained, as initially suspected, by only the variation of a single component 
centered on 20.4\,km\,s$^{-1}$. Observing this red component in the absorption lines of both excited and ground levels, and keeping in mind the measured temperature of this medium by VM17 (1300\,K), 
leads to infer that the dropping component is located at close distance ($\sim$38\,R$_\star$) to the star and varies from a yet unknown process, perhaps exocomets activity drop, as 
proposed by Welsh \& Montgomery (2016) and studied in more details in Section~\ref{sec:exocomets}. 

Since there is no evidence for a blue component in the excited levels lines, it should thus be located at larger distance to the star, certainly several AU where temperatures are 
much lower than 1000\,K. This would explain the absence of a correlation with the inner red component. It is moreover strongly pushed-out in the anti-stellar direction, and 
slowly dissipating. This component might be part of the distant expanding circumstellar gas identified by Brandeker et al. (2004,2011). 

\vspace{-0.5cm}
\section{Exploring the link with \bp\ circumstellar environnement}
\label{sec:exocomets}

\subsection{Exocomets activity in the Ca\,II doublet}

It has been suggested by Welsh \& Montgomery (2016) and VM17 that the circumstellar (CS) Fe\,I originates from the numerous exocomets observed in the system of \bp.~One possible way to 
test this conjecture would be to determine if variations in the Fe\,I absorption lines of the circumstellar gas correspond to variations in the cometary activity, or the quantity  of particles evaporated 
from these exocomets.

In the same HARPS spectra as those used here to analyse Fe\,I lines, there are the Ca\,II doublet lines that were used in Kiefer et al. (2014) to show that the $\beta$\,Pic's exocomets 
separate into 2 families. The D-family would be composed of strongly evaporating comets about a common orbit, while the S-family would be composed of older comets with smaller 
amounts of gas released in the circumstellar medium. The D-family absorption signatures are all located within the -10 to 50\,km\,s$^{-1}$ range in the $\beta$\,Pic rest frame, while 
the S-familly signatures are scattered on a wider velocity range, spanning from -100 to 150 km\,s$^{-1}$.

In order to quantify the cometary activity around $\beta$\,Pictoris, we can calculate the average absorption depth in different velocity domains of the Ca\,II normalized 
spectrum. In the small region about the tip of the circumstellar line that reaches almost zero, there could be deep features strongly blended with the circumstellar line; we thus first exclude 
that region from the analysis, within the range bounded by the instrument resolution ($\pm$2.6\,km\,s$^{-1}$) about 21.57\,km\,s$^{-1}$, the tip of the CS line close to the 
$\beta$\,Pic systemic velocity. We fixed the velocity domain for the D-family to be +5-25\,km\,s$^{-1}$ and for the S-family to be 50-100\,km\,s$^{-1}$. These two domains are not 
overlapping and are centered on the core regions of each family detection statistics, as found in Kiefer et al. (2014).

The reference spectrum that is divided out of the spectra contains the stellar lines and the stable circumstellar and interstellar components. 
Thus, in the normalized spectra, only flux variations due to exocomet absorption or circumstellar disk fluctuations remain.

The average absorption depth (AAD, hereafter) is measured by averaging each normalized spectrum over the specified spectral band. It is thus proportional to the equivalent width, by a factor 
$\Delta\lambda$, the wavelength width of the calculation window. Therefore, as long as the absorbing medium is optically thin, this quantity reflects the amount of materials 
released in the circumstellar medium by transiting exocomets. It also provides information on the typical depth of signatures that are present in the spectrum within the velocity 
bounds. 
\onecolumn
\begin{landscape}
\pagestyle{empty}
\begin{longtable}{@{}lcccccccccc@{}}
\caption{\label{tab:FeI_exc_variation} The temporal variations of the Fe\,I excited levels in $v_\text{Fe\,I~Exc}$, heliocentric velocity (km/s), 
$b_\text{Fe\,I~Exc}$ value (km/s) and Log N column density (in Log~{\rm cm}$^{-2}$, identified by the energy level in cm$^{-1}$) 
evaluated by fitting simultaneously all excited levels absorption lines, epoch by epoch. The sum of the column density in all excited levels is given on the last line. The $\dagger$ sign 
individualizes the rows with the cleanest column density determination. Those are plotted as solid lines in Fig.~\ref{Temporal_variations2}. 
} \\
\hline
 Periods                  &	2003-2004                &	2004-2007              &	2007-2008              &	2008-2009               &	2009-2010               &	2010-2011               &	2013-2014               & 2014-2015                 & 2015-2016                & 2016-2017   \\ 
  Nb. Obs.                & 215                       &	42                     & 	198                & 417                     &  54	                    &	207                     &        453                 &	    60                  &     190                  &  1080           \\ 
 \hline 
$v_\text{Fe\,I~Exc}$             &	20.39$\pm$0.03           &	20.37$\pm$0.09         &	20.39$\pm$0.08         &	20.45$\pm$0.03           &	20.47$\pm$0.14          &	20.33$\pm$0.07          &	20.34$\pm$0.08          &	20.85$\pm$0.17          &  20.19$\pm$0.15          &  20.18$\pm$0.06  \\
$b_\text{Fe\,I~Exc}$             & 	0.80$\pm$0.10            &	0.78$\pm$0.33          &	1.20$\pm$0.20          &	0.98$\pm$0.09            &	1.00$\pm$0.53           &	1.22$\pm$0.17           &	0.89$\pm$0.29           &	1.47$\pm$0.51           &  1.14$\pm$0.54           &   1.00 (fixed$^\dagger$)   \\
\hline
$\dagger$\,\!$\log\,$N$_{\rm FeI~416}$ &	11.61$\pm$0.02           &	11.55$\pm$0.05         &	11.64$\pm$0.03         &	11.60$\pm$0.02           &	11.54$\pm$0.07          &	11.65$\pm$0.03          &	11.39$\pm$0.04          &	11.54$\pm$0.08          &  11.31$\pm$0.08          &   11.45$\pm$0.03               \\
$\log\,$N$_{\rm FeI~704}$ &	11.25$\pm$0.03           &	10.84$^{+0.15}_{-0.28}$&	11.38$\pm$0.05         &	11.40$\pm$0.03           &	11.15$^{+0.15}_{-0.24}$ &	11.30$\pm$0.08          &	11.09$\pm$0.08          &	11.11$\pm$0.15          &  10.84$^{+0.16}_{-0.36}$ &   11.16$\pm$0.06               \\
$\log\,$N$_{\rm FeI~888}$ &	11.01$\pm$0.06           &	11.18$\pm$0.08         &	10.69$^{+0.20}_{-0.30}$&	10.88$\pm$0.07           &	$<$10.96                &	11.03$\pm$0.10          &	10.81$^{+0.11}_{-0.16}$ &	10.86$^{+0.20}_{-0.37}$ & 10.73$^{+0.20}_{-0.50}$  &   10.74$^{+0.12}_{-0.20}$      \\
$\log\,$N$_{\rm FeI~978}$ &	10.51$^{+0.08}_{-0.12}$  &	$<$10.47               &	$<$10.10               &	10.45$^{+0.16}_{-0.29}$  &	10.73$^{+0.24}_{-0.57}$ &	$<$10.65                &	10.50$^{+0.12}_{-0.19}$ &	$<$10.65                &   $<$10.06               &   $<$9.91                       \\
\hline
$\dagger$\,\!$\log\,$N$_{\rm FeI~6928}$ &	10.92$\pm$0.01           &	10.87$\pm$0.04         &	10.88$\pm$0.04         &	10.93$\pm$0.02           &	10.90$\pm$0.07          &	11.03$\pm$0.03          &	10.60$\pm$0.05          &	10.75$\pm$0.08          &   10.67$\pm$0.07           &   10.75$\pm$0.03      \\
$\log\,$N$_{\rm FeI~7377}$ &	10.47$\pm$0.03           &	10.49$\pm$0.08         &	10.45$\pm$0.09         &	10.54$\pm$0.03           &	10.51$^{+0.13}_{-0.19}$ &	10.45$\pm$0.09          &	10.06$^{+0.10}_{-0.14}$ &	9.92$^{+0.31}_{-0.97}$  &   10.22$\pm$0.20           &   10.34$\pm$0.06      \\
$\log\,$N$_{\rm FeI~7728}$ &	$<$9.67                  &	$<$10.74               &	10.83$^{+0.17}_{-0.26}$&	10.27$^{+0.22}_{-0.38}$  &	10.88$^{+0.19}_{-0.33}$ &	$<$10.41                &	$<$9.80                 &	$<$10.23                &   $<$10.40                 &   $<$9.93            \\
$\log\,$N$_{\rm FeI~7986}$ &	$<$9.90                  &	$<$9.98                &	$<$10.14               &	9.87$^{+0.18}_{-0.33}$   &	$<$10.07                &	10.31$^{+0.15}_{-0.24}$ &	10.15$^{+0.16}_{-0.26}$ &	$<$9.67                 &   $<$10.02                 &   $<$9.10               \\
$\log\,$N$_{\rm FeI~8155}$ &	$<$8.90                  &	$<$10.15               &	9.95$^{+0.25}_{-0.78}$ &	10.18$^{+0.10}_{-0.13}$  &	$<$10.36                &	$<$10.08                &	$<$9.35                 &	$<$9.33                 &   10.17$^{+0.18}_{-0.32}$  &   $<$9.63           \\
\hline
$\log\,$N$_{\rm FeI~11976}$ & 9.68$^{+0.04}_{-0.05}$  &	9.56$^{+0.12}_{-0.16}$ &	9.59$^{+0.10}_{-0.12}$ & 9.61$^{+0.07}_{-0.08}$  &	9.55$^{+0.21}_{-0.46}$  &	9.82$\pm$0.08           &	9.43$^{+0.10}_{-0.15}$  &	9.67$\pm$0.13           &    $<$9.11                 &   9.49$\pm$0.12      \\
$\log\,$N$_{\rm FeI~12561}$ & 9.24$^{+0.07}_{-0.11}$  &	$<$8.39                &	9.15$^{+0.24}_{-0.40}$ & 9.35$^{+0.08}_{-0.10}$  &	$<$9.48                 &	9.43$^{+0.14}_{-0.22}$  &	9.00$^{+0.19}_{-0.41}$  &	$<$9.18                 &    9.14$^{+0.23}_{-0.46}$  &   9.12$^{+0.11}_{-0.43}$\\
$\log\,$N$_{\rm FeI~12969}$ & 9.12$^{+0.08}_{-0.11}$  &	9.27$^{+0.14}_{-0.21}$ &	9.35$^{+0.11}_{-0.16}$ & 9.28$^{+0.06}_{-0.09}$  &   $<$9.44                 &	9.24$^{+0.16}_{-0.28}$  &	$<$8.47                 &	9.28$^{+0.19}_{-0.34}$  &    $<$9.11                 &   9.23$^{+0.14}_{-0.20}$\\
\hline
$\dagger$\,\!$\log\,$N$^{\rm Total}_{\rm FeI~Exc}$ &	11.93$\pm$0.01 & 11.84$\pm$0.04 &  11.96$\pm$0.03  &	11.97$\pm$0.02 &	11.87$\pm$0.06 &	11.97$\pm$0.03 &	11.73$\pm$0.03  &	11.80$\pm$0.06 & 11.61$\pm$0.07 & 11.75$\pm$0.03 \\
\end{longtable}
\flushleft
$^\dagger$ The b-value was fixed at the median value from past measurements, because \verb+Owens.f+ was unable to converge on a realistic value.
\begin{longtable}{@{}lcccccccccc@{}}
\caption{\label{tab:ground_2comp} Table of fitting parameters, modelizing the Fe\,I ground level absorption lines with 2 components.} \\
\hline
 Periods                               &	2003-2004  			&	2004-2007            	     &	2007-2008           				&	2008-2009          		     &	2009-2010      	         &	 2010-2011               &	2013-2014               & 2014-2015       	       & 2015-2016               & 2016-2017   \\ 
  Nb. Obs.                             & 215       			 &	42                   	     & 	198               			    & 417                  			 &  54	                     &	207                      &        453                &	    60            	   &     190                 &  1080           \\ 
\multicolumn{11}{l}{\textit{Comp 2 fixed to v=20.39\,km\,s$^{-1}$}} \\ \\
$V_\text{comp,1, {\rm FeI~0}}$         & 19.48$^{+0.11}_{-0.05}$  & 20.08$^{+0.05}_{-0.11}$      & 19.72$\pm$0.20           	    & 19.73$^{+0.15}_{-0.06}$        &  19.84$^{+0.15}_{-0.40}$  &    19.81$^{+0.04}_{-0.17}$   &  19.48$\pm$0.10           &  19.96$\pm$0.11           & 19.85$\pm$0.20          &  19.78$^{+0.20}_{-0.11}$ \\
$b_\text{comp,1, {\rm FeI~0}}$         &  0.04$^{+0.02}_{-0.01}$  & 0.16$\pm$0.05                & $<$0.07                 	    &  0.06$\pm$0.02                 &   $<$0.09                 &    $<$0.06                   &  $<$0.05                  &   0.09$^{+0.06}_{-0.04}$  &  $<$0.14                &   0.20$^{+0.23}_{-0.14}$ \\
$b_\text{comp,2, {\rm FeI~0}}$         &  0.55$^{+0.08}_{-0.04}$  & 1.50$^{+0.55}_{-0.28}$       & 0.87$^{+0.38}_{-0.10}$    	    &  0.77$\pm$0.03                 &   0.65$^{+0.48}_{-0.20}$  &    0.99$\pm$0.08             &  0.76$^{+0.05}_{-0.08}$   &   1.17$^{+0.40}_{-0.20}$  &  0.43$^{+0.17}_{-0.13}$ &   0.19$^{+0.09}_{-0.03}$ \\
$\log\,$N$_\text{comp 1, {\rm FeI~0}}$ & 12.02$\pm$0.03           & 12.36$\pm$0.06               & 12.07$^{+0.06}_{-0.12}$  	    &  12.01$\pm$0.04                &  12.20$^{+0.09}_{-0.30}$  &    12.03$^{+0.07}_{-0.12}$   &  11.86$\pm$0.07           &  12.04$^{+0.05}_{-0.15}$  & 11.95$^{+0.05}_{-0.15}$ &  11.79$\pm$0.09 \\
$\log\,$N$_\text{comp 2, {\rm FeI~0}}$ & 12.28$^{+0.01}_{-0.03}$  & 11.94$^{+0.15}_{-0.11}$      & 12.24$^{+0.03}_{-0.16}$    	    &  12.29$^{+0.01}_{-0.06}$       &  12.25$^{+0.02}_{-0.08}$  &    12.29$\pm$0.02            &  12.12$\pm$0.02           &  12.00$^{+0.07}_{-0.14}$  & 11.96$\pm$0.10          &  12.20$^{+0.04}_{-0.11}$\\
$\chi^2$   (DOF=137)                   &  142.86                  & 143.03               	       &  142.63                 	    &   154.14                       &   139.22                  &    140.21                    &    144.99                 &   142.70                  &   138.39                &   137.85 \\
\hline
\hline
\multicolumn{11}{l}{\textit{Comp. 1 fixed to v=19.77\,km\,s$^{-1}$}} \\ 
\multicolumn{11}{l}{\textit{Comp. 2 fixed to v=20.39\,km\,s$^{-1}$}} \\ \\
$b_\text{comp,1, {\rm FeI~0}}$         & 0.14$\pm$0.02            & 0.06$\pm$0.02                & 0.05$\pm$0.03                &  0.08$\pm$0.02                   &  $<$0.04                    &    $<$0.04                 &  0.08$^{+0.04}_{-0.02}$     &  $<$0.12                    & 0.06$^{+0.06}_{-0.03}$      &  0.28$\pm$0.15\\
$b_\text{comp,2, {\rm FeI~0}}$         & 0.81$\pm$0.03            & 0.90$\pm$0.08                & 0.91$\pm$0.05                &  0.82$\pm$0.03                   &  0.63$\pm$0.13              &    1.00$\pm$0.05           &  1.01$\pm$0.05              &  0.98$\pm$0.10              & 0.43$\pm$0.15               &  0.20$^{+0.09}_{-0.03}$\\
$\log\,$N$_\text{comp 1, {\rm FeI~0}}$ & 12.16$\pm$0.02           & 12.19$\pm$0.05               & 12.09$\pm$0.06               &  12.05$\pm$0.03                  &  12.19$\pm$0.10             &    12.04$\pm$0.06          &  11.99$\pm$0.04             &  11.94$\pm$0.09             &  11.92$\pm$0.06             &  11.82$\pm$0.03\\
$\log\,$N$_\text{comp 2, {\rm FeI~0}}$ & 12.18$\pm$0.01           & 12.21$\pm$0.02               & 12.23$\pm$0.02               &  12.27$\pm$0.01                  &  12.25$\pm$0.03             &    12.29$\pm$0.02          &  12.04$\pm$0.02             &  12.08$\pm$0.03             & 11.99$\pm$0.03              &  12.19$\pm$0.03\\
$\chi^2$  (DOF=138)                    &  153.79        		     & 147.04 	                   &  143.11                      &   156.63                         &   139.24                    &    140.23                  &    149.60                   &   143.93                    &   138.39                    &   137.87 \\
\hline
\end{longtable}
\end{landscape}
\twocolumn

\begin{figure}
\includegraphics[width=89mm]{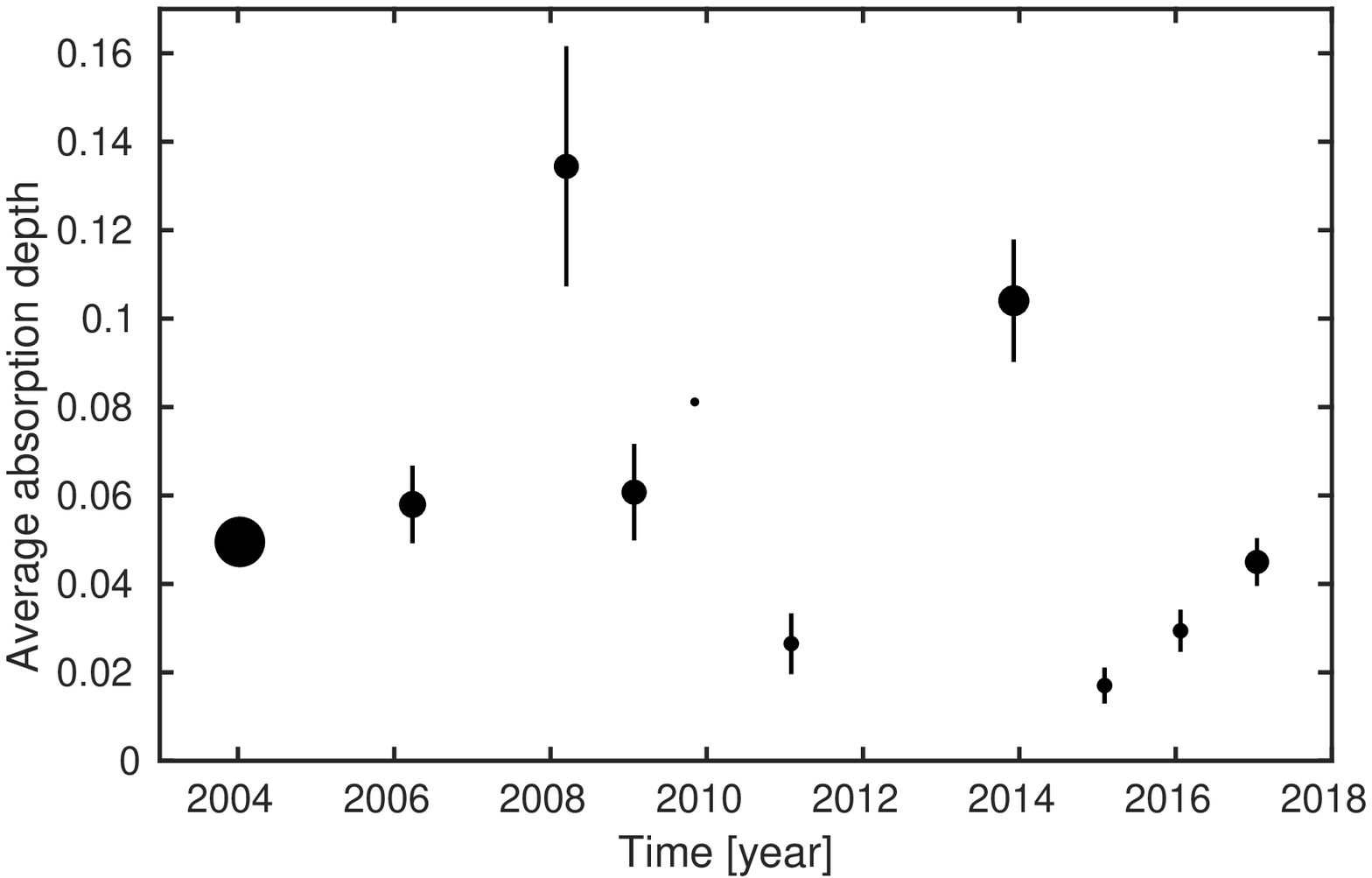}
\includegraphics[width=89mm]{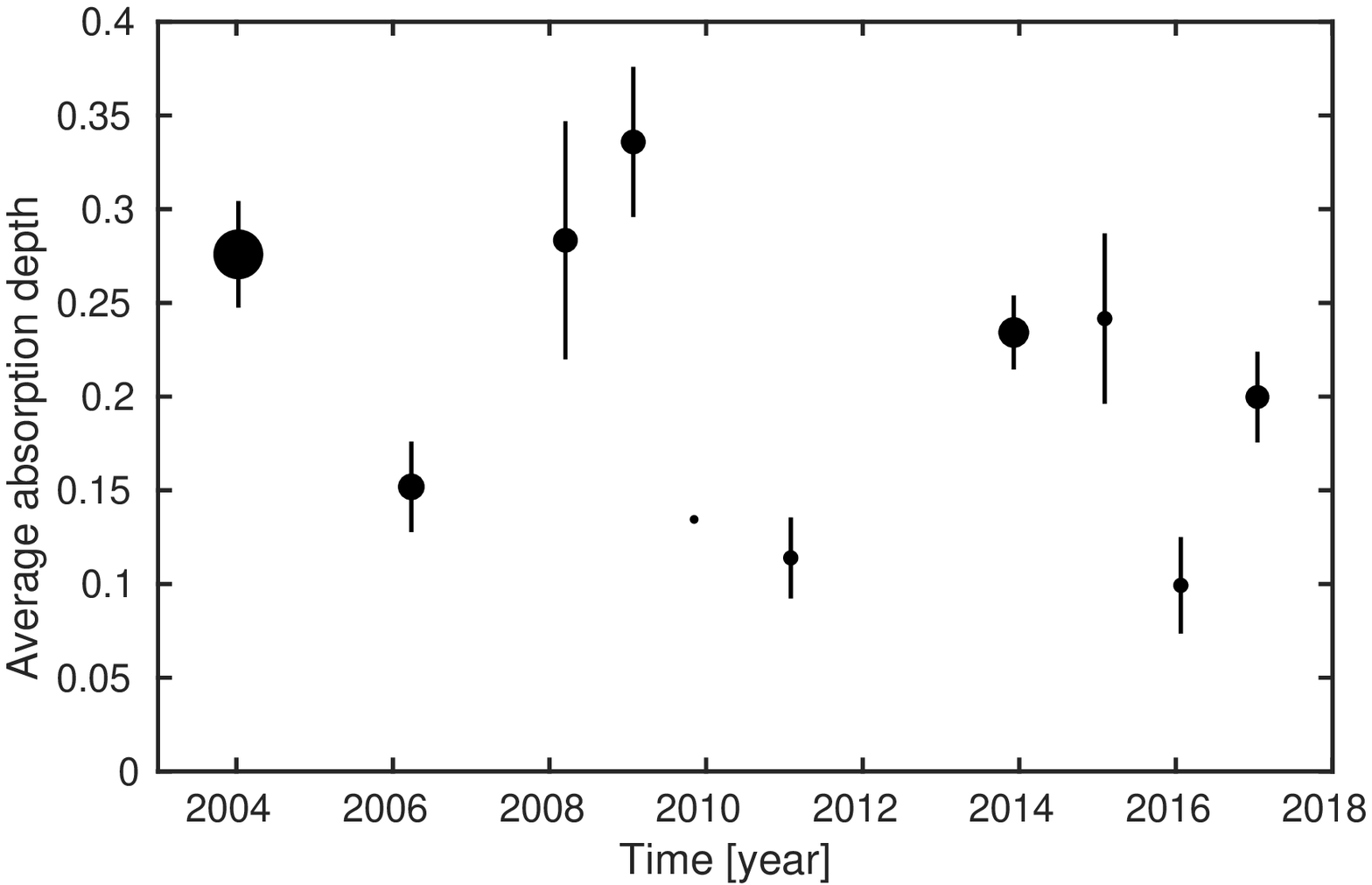}
\caption{\label{fig:eqW} Average absorption depth variations in the co-added Ca\,II K and H absorption spectra. \textbf{ Top:} S-familly, between 50 and 100\,km\,s$^{-1}$. 
\textbf{Bottom:} D-familly, between 5 and 25\,km\,s$^{-1}$. The marker size is proportionnal to the number of nights where $\beta$ Pic was observed by HARPS in each period 
(see Table~\ref{tab:repartition}). The errorbars indicate the scatter of average absorption depth during each period. See text for explanations.}
\end{figure}

Most of the time, several spectra are observed on the same night; this allows us to obtain an average AAD per night, a measure of exocometary activity during a single night. 
Averaging these night-based AAD over each of the periods considered in this paper (Table 1) we get a measure of how much the exocometary activity varies through time from one 
period to the other. It could then be compared to the Fe\,I variations in Fig.~\ref{fig:ground_2c}. Using the night-based AAD for this computation, rather than averaging over all spectra of a 
given period, makes more sense, because it prevents nights with many collected spectra to dominate the period-based AAD. These average absorption depth are plotted for each 
exocomet familly on Figure~\ref{fig:eqW}.

As can be seen, there is no obvious long-term variations or sudden drop in the exocomet activity. On the contrary, the measurements scatter and the 
errorbars show that the total quantity of particles evaporated from the many transiting exocomets is variable even within a given period. 
Thus we see no evidence here for any correlation between exocomet activity and Fe\,I disk column densities.

\subsection{Variability in the core of the Ca\,II CS line}
\begin{figure}
\includegraphics[width=89mm]{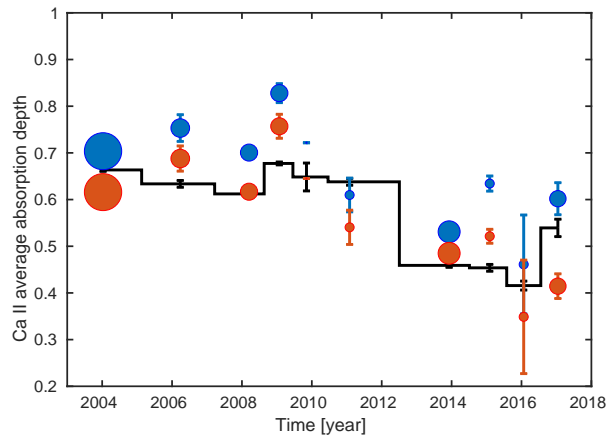}
\caption{\label{fig:AAD} Average absorption depth (AAD) variations in the Ca\,II circumstellar line region 0$\pm$5\,km\,s$^{-1}$. In blue, the K-line AAD and in red the 
H-line. They are compared to the total Fe\,I ground level column density variations in Table~\ref{tab:FeIvariation}. The marker size is proportionnal to the number of 
nights where $\beta$ Pic was observed by HARPS in each period (see Table~\ref{tab:repartition}). For the sake of comparison, the Fe\,I column densities are scaled to the 
Ca\,II AAD median.
}
\end{figure}

We first excluded the strongly blended CS line region (0$\pm$5\,km\,s$^{-1}$), since we were initially interested in exocomet absorption. However, the CS line region 
incorporates circumstellar disk variations, as well as exocomets transit signatures. It is interesting to see if significant Ca\,II variations show up in that region, even though 
we are not able to know their origin. We plotted the AAD variation in this domain in Fig.~\ref{fig:AAD}. 

This time, we observe a neat decrease in average absorption depth. The varying blended 
features in that region have an average depth of about 0.68$\pm$0.03 in 2003-2011 and decrease to 0.50$\pm$0.04 in 2013-2017, for a total 3.6$\sigma$ drop of 26\% along the 14 
years of observations. This is comparable to the measured drop of column density of Fe\,I. This variation is confirmed by measuring a flux increase in the 5 pixels at the tip of the K-line from 0.0096$\pm$0.0012 to 
0.0196$\pm$0.0021 in arbitrary unit, and from 0.0166$\pm$0.0019 to 0.037$\pm$0.0064 at the tip of the H-line. 

Given that the average absorption depth is directly proportional to equivalent width, and thus at first approximation to average column density in the absorbing medium, we see that 
both Ca\,II medium and Fe\,I medium varied in about the same proportions. This has implications on the Fe and Ca relative abundance in the circumstellar medium close to the star, that 
we will explore in more details in the next section.

In conclusion, the variations experienced by the Fe\,I components column density are likely connected to Ca\,II variations in the circumstellar medium. Ca and Fe are thus
part of a common reservoir that suddenly dissipated between 2011 and 2013. We cannot firmly identify the origin of these variations, which could be either large low-velocity 
exocomets or local gas disk inhomogeneities. The exocomets scenario is however less likely as they would have to be disconnected to the already known exocomets which
families show no sign of long-term or sudden variations.
 
\begin{table*}[t]
\caption{The variation of the Ca\,II average absorption depth and average flux around Ca\,II K and H circumstellar lines. Average absorption depths are calculated from the 
normalized spectra, either separating K and H spectra, either co-adding the K and H absorption lines. 
Average fluxes are calculated from the raw spectra. Their continuum is scaled to an arbitrary level that is common to all spectra.}
\label{tab:CaIIvariation}
\begin{small}
\begin{tabular}{@{}c@{~~}|c@{~~}c@{~~}c@{~~}c@{~~}|c@{~~}c@{~~}c@{}}
\hline
Periods	& \multicolumn{4}{c|}{Average absorption depth}    													&  \multicolumn{3}{c}{Average flux [arb. unit]}\\ 
	 	&	      \multicolumn{2}{c}{CS-line}										&	D-family 		&	S-family 									&	\multicolumn{2}{c}{tip of CS-line}							& blue wing   \\		
		&  K only          & H only	       	           									& \multicolumn{2}{c|}{K \& H coadded}					&		K	only						& 			H	only					&   K only	\\
		&  $\pm$5\,km\,s$^{-1}$ & $\pm$5\,km\,s$^{-1}$         		& 5 - 25\,km\,s$^{-1}$ 	& 50 - 100\,km\,s$^{-1}$	&	0$\pm$2\,km\,s$^{-1}$	& 0$\pm$2\,km\,s$^{-1}$	&	-150$\pm$5\,km\,s$^{-1}$\\
\hline
2003-2004 & 0.704$\pm$0.014 	&	0.61597$\pm$0.010	& 0.276$\pm$0.028 	& 0.0495$\pm$0.0045	&	0.01118$\pm$0.00063	&	0.01843$\pm$0.00062	& 0.3881$\pm$0.0010\\
2004-2007 & 0.753$\pm$0.029 	&	0.68788$\pm$0.027	& 0.152$\pm$0.024 	& 0.0580$\pm$0.0088	&    0.0105$\pm$0.0016   & 0.0162$\pm$0.0024	&	0.3864$\pm$0.0024\\
2007-2008 & 0.701$\pm$0.013 	&	0.61683$\pm$0.013	& 0.283$\pm$0.064 & 0.134$\pm$0.027		&    0.01146$\pm$0.00074     & 0.01679$\pm$0.00080	&	0.3768$\pm$0.0038\\
2008-2009 & 0.828$\pm$0.020 	&	0.75697$\pm$0.026	& 0.336$\pm$0.040	& 0.061$\pm$0.011		&   0.00437$\pm$0.00059 	& 0.00801$\pm$0.00091 	&	0.3810$\pm$0.0017 \\
$^\dagger$2009-2010& 0.72174 	&	0.64518 					& 0.135						& 0.081							&		 0.0079  &		0.0204 	&	0.3869\\
2010-2011 & 0.610$\pm$0.036		&	0.54054$\pm$0.037& 0.114$\pm$0.022	& 0.0265$\pm$0.0069	& 0.0124$\pm$0.0022	&	0.0198$\pm$0.0029 	&	0.39563$\pm$0.00057 \\
2013-2014 & 0.531$\pm$0.012		&	0.48492$\pm$0.012	& 0.234$\pm$0.020	& 0.104$\pm$0.014		&   0.02185$\pm$0.00060 	& 0.02788$\pm$0.00065	&	0.3820$\pm$0.0024\\
2014-2015 & 0.634$\pm$0.016 	&	0.52135$\pm$0.015	& 0.242$\pm$0.045	& 0.0170$\pm$0.0041	&  0.01436$\pm$0.00060 &	0.0233$\pm$0.0011	&	0.3949$\pm$0.0014\\
2015-2016 & 0.46$\pm$0.11			&	0.34875$\pm$0.12	& 0.099$\pm$0.035	& 0.0294$\pm$0.0048	&  0.0241$\pm$0.0060 &	0.049$\pm$0.012	& 0.3950$\pm$0.0021\\
2016-2017 & 0.602$\pm$0.034 	&	0.4145$\pm$0.026	& 0.200$\pm$0.024	& 0.0450$\pm$0.0054	&  0.0182$\pm$0.0012 &	0.04530$\pm$0.00092	& 0.3858$\pm$0.0013\\
\hline
\end{tabular}
\end{small}
$^\dagger$With only 1 night observed it is not possible to evaluate uncertainties.
\end{table*}

\section{Discussing the relation between Ca and Fe} 
\subsection{Implication for the Fe\,II/Fe\,I ionization ratio}
Independently in both K and H line of the Ca\,II doublet, we measure an average absorption depth drop of about 0.36$\pm$0.04. The average K/H-line 
ratio of the dropped Ca\,II component is therefore about 1$\pm$0.2. 
Comparing the individuals AAD of K and H lines in Table~\ref{tab:CaIIvariation} leads to a refined K/H ratio closer to 1.2.
Thus, the Ca\,II varying medium is likely \textit{not fully} saturated. 
In terms of equivalent width, with the core of the variation concentrated within $\pm$5\,km\,s$^{1}$ of \bp\ velocity, we can estimate that 
\begin{equation}
W_{K,H} = AAD\times\frac{10\text{\,km\,s}^{-1}}{c}\times \lambda_{K,H}
\end{equation}

The lower limit from simple linear relation between equivalent width and column density in the K and H lines gives that the variation of the column density of 
Ca\,II should be $\Delta N_\text{Ca\,II}$$>$$4.5\,10^{11}$ cm$^{-2}$. 
Using Somerville (1988) equivalent width ratio method, best to use in close-to-saturated cases, we find a range of possible column density for a ratio of 1.2:
\begin{equation}
\Delta N_\text{Ca\,II}\sim 1-4\,10^{12}\,\text{cm}^{-2}
\end{equation}

This should be compared to the $\Delta N_\text{Fe\,I}$=$(7.5$$\pm $$0.9)\,10^{11}$\,cm$^{-2}$ lost during the 2011-2013 drop in the ground level column density. Assuming that Ca is fully 
ionized below 1\,au (Fernandez et al. 2006), we can obtain an estimation of the abundance of Ca with respect to Fe in this medium, by calculating 
$\Delta N_\text{Ca\,II}/\Delta N_\text{Fe\,I}$. We find a ratio of about 
\begin{equation}
\text{Ca\,II}/\text{Fe\,I}\sim 1-5
\end{equation}

Therefore, Ca\,II and Fe\,I are almost as abundant in this medium. If it follows $\beta$\,Pic standard abundances (Roberge et al. 2006), with 
Fe/Ca$\sim$15, then we must have Fe\,II/Fe\,I$\sim$15-75, implying a low ionization rate for Fe. This is in fairly good agreement with the results of VM17, proposing that 
Fe\,I/Fe\,II$\lesssim$1 in the 20.4\,km\,s$^{-1}$ component at 1300\,K.

\subsection{Ca\,I line variations}
\begin{figure}
\includegraphics[height=89mm, angle=90]{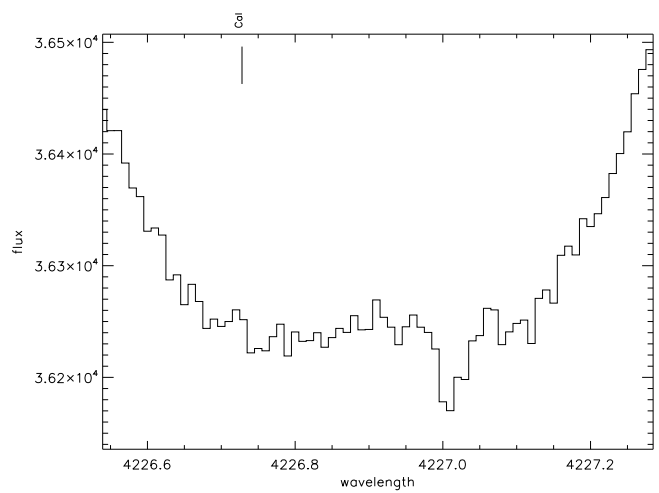}
\caption{\label{fig:CaI} Ca\,I absorption line at 4226.728\,\AA~with a shift of 20\,km\,s$^{-1}$. All HARPS spectra from 2003 to 2017 are here co-added.}
\end{figure}

\begin{figure}
\includegraphics[width=89mm]{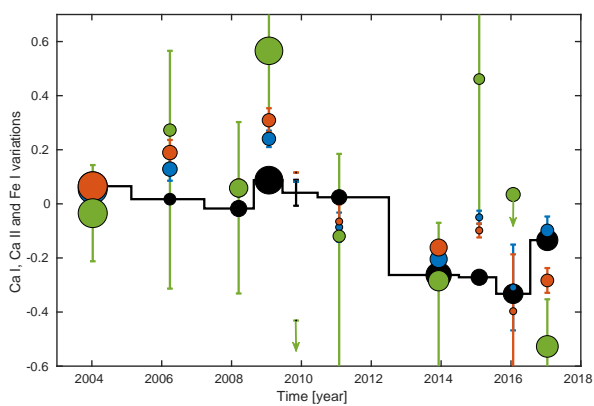}
\caption{\label{fig:caI_caII_feI} Scaled-to-median Ca\,I column density variations at 4226.728\,\AA, in green, compared to Ca\,II-K (in blue), Ca\,II-H (in red) and Fe\,I variations (black). 
Absolute values for Ca\,I column densities are reported in Table~\ref{tab:CaI_var}. The median of all datasets was shifted to 0. The size is proportional to the number
of observed nights for Ca\,II, while it is proportional to SNR for Fe\,I and Ca\,I, as given in Tables~\ref{tab:CaIIvariation},~\ref{tab:FeIvariation} and~\ref{tab:CaI_var} respectively.}
\end{figure}

In the HARPS spectra, we also found a Ca\,I circumstellar absorption line about 4226.728\,\AA, as plotted on Fig.~\ref{fig:CaI}. Measuring its variation allow comparing Ca\,I and Fe\,I and determining 
independently the Ca/Fe abundance ratio. We thus analysed as for Fe\,I the variation of the column density in this line using \verb+Owens.f+. The results are given in 
Table~\ref{tab:CaI_var}, and compared to Fe\,I and Ca\,II variations in Fig.~\ref{fig:caI_caII_feI}.

Again we found a 2-$\sigma$ significant drop of column density compatible with 30\% between 2011 and 2013. The measured column density variation is $\Delta N_\text{Ca\,I}$=$(1.54$$\pm $$0.70)\times10^8$\,cm$^{-1}$.
This means that Ca\,I, Ca\,II and Fe\,I variations are all compatible and most likely originate from a common medium.

First comparing the column density variation of Ca\,I, to the one estimated above for Ca\,II, we find that indeed 
Ca is almost fully ionized with a ratio Ca\,II/Ca\,I$\sim$10$^4$, much higher than what primarily found by Hobbs et al. (1985) but closer to the theoretical estimations of Fernandez et al. (2006).
 Second, comparing this Ca\,I variation to $\Delta N_\text{Fe\,I}$ we find that Fe\,I is $\sim$5000 ($\pm$2400) times more abundant than Ca\,I. If originating from a common 
 medium, then according to photoionisation-recombination balance (Lagrange et al. 1995), Ca and Fe should follow, at a given distance to the central star, a one-to-one correspondance in first ionization ratio
\begin{equation}
N_\text{Fe\,I}/N_\text{Fe\,II} = 200 \times N_\text{Ca\,I}/N_\text{Ca\,II}
\end{equation}

 Assuming that $N_\text{Fe}$$\sim $$N_\text{Fe\,II}$ and $N_\text{Ca}$$\sim $$N_\text{Ca\,II}$ is a fairly good approximation even if Fe\,II/Fe\,I$\sim$10. In this case, we derive an 
 abundance ratio Fe/Ca of about 25 ($\pm$12). It is compatible with the solar value for this ratio $\sim$15 (Lodders 2003, Lodders 2010) and the value found in $\beta$\,Pic by Roberge et al. (2006), 
 15$\pm$10. 
 
 We can also confirm that Fe ionisation ratio is low with the direct measurement of Fe\,II/Fe\,I$\sim$30-100, using the above formula and the estimation 
 of $N_\text{Ca\,I}/N_\text{Ca\,II}$.

\begin{table}
\caption{The variation of the Ca\,I velocity shift, b value and column density evaluated by fitting the ground base line at 4226.728\,\AA. Upper-limits for b-values are given at 1-sigma.}
\label{tab:CaI_var}
\begin{tabular}{@{}ccccc@{}}
\hline
Periods   & SNR 	&    v            			& b              	& $\log$N$_{\rm FeI}$  \\ 
 (year)   &   	&  (km\,s$^{-1}$)         			& (km\,s$^{-1}$)         	&    (cm$^{-2}$)      \\
\hline
2003-2004 & 2120	&  19.80$\pm$0.25 			&	$<$2.0 			&	8.48$\pm$0.08 \\
2004-2007 & 870	&  21.29$\pm$0.42 			&	$<$1.4	 		&	8.60$^{+0.10}_{-0.20}$ \\
2007-2008 & 1200	& 19.53$\pm$0.35 			&	$<$1.00 			&	8.52$^{+0.10}_{-0.16}$ \\
2008-2009 & 1980	& 19.85$\pm$0.22 			&	1.30$\pm$0.45 	&	8.69$\pm$0.09 \\
2009-2010 & 590	& 15.42$^{+0.93}_{-0.67}$ 	&	--- 				&	$<$8.25 $^\dagger$\\
2010-2011 & 860	& 19.56$\pm$0.63 			&	$<$1.5 			&	8.44$^{+0.15}_{-0.43}$ \\
2013-2014 & 1350	& 20.08$\pm$0.45 			&	$<$1.1  			&	8.35$^{+0.13}_{-0.25}$ \\
2014-2015 & 800	& 20.70$\pm$0.44 			&	$<$1.6 			&	8.66$\pm$0.15 \\
2015-2016 & 950	& 31.40$^{+0.35}_{-0.70}$	&	---				&	$<$8.51 $^\dagger$ \\
2016-2017 & 1450	& 20.35$\pm$0.65 			&	$<$1.1 			&	8.17$^{+0.16}_{-0.63}$ \\
\hline
\end{tabular}
$^\dagger$2-sigma upper-limit calculated varying $v$ in range 19-21\,km\,s$^{-1}$. 
\end{table}

\section{Conclusion}

To summarize the reported results, we observed in the circumstellar gas disk of \bp\ that 
\begin{itemize}
\item The Fe\,I ground level column density drop of 2011-2013 is also observed in the Fe\,I excited levels absorption lines centered around velocity 20.4\,km\,s$^{-1}$,
\item The blue and red components of Fe\,I$_0$ absorption lines have different variability, with the 20.4\,km\,s$^{-1}$ more compatible with a sudden drop, while the blue component
seems to have an independent behavior,
\item We identified a varying Ca\,II component in the circumstellar line region, which equivalent width is dropping in average with an amplitude comparable to that of Fe\,I.
\item The Ca\,I circumstellar line also experiences a drop between 2011 and 2013. This drop is compatible with $\beta$\,Pic-like relative abundances for Ca and Fe,
\item The varying component of Fe has a low ionisation ratio, in agreement with VM17 results. 
\end{itemize} 

First we conclude that the VM17 1300\,K medium at 20.4\,km\,s$^{-1}$ contains not only Fe, but also Ca, and both stands at photoelectric equilibrium. It is located at close 
distance to the star $\sim$38\,$R_\star$ to sustain the 1300\,K temperature and the high ionisation ratio of Ca. The Fe\,I blue component is likely not connected to this inner 
disk and belongs to a colder outer location with different photoionization conditions, possibly from recombination of Fe\,II beyond 100\,AU, as described by Brandeker et al. (2004,2011). 

Second, although the Ca\,II absorptions due to exocomets in the \bp\ spectrum have a strong variability, they do not show any long term or sudden variations as in the 
column density of the 20.4\,km\,s$^{-1}$ component of Fe\,I. 
On the other hand, depth variations in the core of the Ca\,II circumstellar line have a stronger compatibility with the Fe\,I drop. Large and slow transiting exocomets could be at the origin of such 
absorptions, blended within the circumstellar line, such as members of the D-family, as suggested by Welsh \& Montgomery (2016). However, the absence of correlated variations at larger velocity
(up to +30\,km\,s$^{-1}$ in $\beta$\,Pic rest frame) where most D-family objects lie does not support this hypothesis. 

Small scale disk inhomogeneities with yearly
density variations are another alternative scenario. Such variations could be triggered by an outer planet, through direct gravitational interaction or indirectly by changing the flux of 
incoming dust within an inner disk at 0.3\,au. However in that case a periodic behaviour would be expected. The baseline is not long enough to confirm or exclude periodicity, but 
a slight increase of the Fe\,I column density is already observed. The continuation of the monitoring of $\beta$\,Pic optical spectrum with high resolution spectrograph
such as HARPS will allow testing this scenario.

\begin{acknowledgements}
We thank the anonymous referee for his help on greatly improving the quality of the article. We warmly thank T.~Lanz for fruitful discussions on the subject of 
the present article. F.K. acknowledge support by a CNES fellowship grant. A.L.E, A.V-M and F.K thank the CNES for financial support. This work has been partly carried out 
thanks to an award from the Fondation Simone et Cino Del Duca. P.A.W. acknowledge support from the European Research Council under the European Unions Horizon 2020 
research and innovation programme under grant agreement No. 694513. V. B. and D. E. acknowledge support by the Swiss National Science Foundation (SNSF) in the frame of the National 
Centre for Competence in Research PlanetS, and has received funding from the European Research Council (ERC) under the European Union's Horizon 2020 research and 
innovation programme (project Four Aces; grant agreement No 724427). HARPS data were obtained at ESO 3.6m telescope from 2003 to 2017, with Program IDs, 60.A-9036, 072.C-0636, 
075.C-0234, 076.C-0073, 076.C-0279, 078.C-0209, 079.C-0170, 080.C-0032, 080.C-0664, 080.C-0712, 081.C-0034, 082.C-0308, 082.C-0412, 084.C-1039, 091.C-0456, 094.C-0946, 
098.C-0739, 099.C-0205, 099.C-0599, 184.C-0815, \& 192.C-0224

\end{acknowledgements}

\end{document}